# Title: Risk of SARS-CoV-2 in a car cabin assessed through 3D CFD simulations

**Authors:** Fausto Arpino[1], Giorgio Grossi[1], Gino Cortellessa[1], Alex Mikszewski[2], Lidia Morawska[2], Giorgio Buonanno[1,2], Luca Stabile[1]

**Affiliations:**
[1] Department of Civil and Mechanical Engineering, University of Cassino and Southern Lazio, Cassino, FR, Italy
[2] International Laboratory for Air Quality and Health, Queensland University of Technology, Brisbane, Queensland, Australia

**Abstract**

In this paper the risk of infection from SARS-CoV-2 Delta variant of passengers sharing a car cabin with an infected subject for a 30-min journey is estimated through an integrated approach combining a recently developed predictive emission-to-risk approach and a validated CFD numerical model numerically solved using the open-source OpenFOAM software. Different scenarios were investigated to evaluate the effect of the infected subject position within the car cabin, the air flow rate of the HVAC system, the HVAC ventilation mode, and the expiratory activity (breathing vs. speaking).

The numerical simulations here performed reveal that the risk of infection is strongly influenced by several key parameter: as an example, under the same ventilation mode and emitting scenario, the risk of infection ranges from zero to roughly 50% as a function of the HVAC flow rate. The results obtained also demonstrate that: (i) simplified zero-dimensional approaches limit proper evaluation of the risk in such confined spaces, conversely, (ii) CFD approaches are needed to investigate the complex fluid-dynamics in similar indoor environments, and, thus, (iii) the risk of infection in indoor environments characterized by fixed seats can be in principle controlled by properly designing the flow patterns of the environment.

**Practical Implications**

The study investigated the risk of infection in car cabins combing a predictive emission-to-risk approach and a validated CFD approach. The findings illustrate a methodology for designing proper ventilation systems for car cabin in view of reducing and controlling the risk of infection of passengers.

**Conflict of Interest:** The authors have no conflicts of interest to declare.

## 1 Introduction

Transport microenvironments are confined spaces of concern in terms of SARS-CoV-2 risk of infection due to the high crowding indexes (number of people relative to the size of the confined space) and the possible inadequate clean (pathogen-free) air supply. Indeed, a number of outbreaks occurred worldwide in buses, airplanes, and ships [1–8]. These outbreaks are mainly due to the airborne transmission of inhalable virus-laden airborne respiratory particles (i.e. particles below 100 μm in diameter) which are capable of remaining suspended in the air then likely infecting simultaneously numerous susceptible subjects that share the same confined space of the infected subject. This route of transmission was accepted as the main pathway of infection transmission only in the spring 2021 when, faced with the accumulating scientific evidence [9–15], both US CDC and WHO released updated guidelines reporting the negligible role of the sprayborne particles (larger particles >100 μm quickly settling due to their inertia) and fomites (i.e. contaminated surfaces) with respect to the airborne respiratory particles (WHO, April 30th, 2021; US CDC, May 7th, 2021). Indeed, SARS-CoV-2 virus has been detected in airborne samples collected in indoor environments such as hospital microenvironments [12–14,16] (where certain presence of infected subjects allows simpler particle samplings) but also in transport microenvironments [17,18], including passenger cars [19]. Thus, despite the implementation of sprayborne and surface touch mitigation strategies (i.e. wearing cloth or surgical masks, washing hands), whose effectiveness on airborne respiratory particles is questionable, reducing the airborne SARS-CoV-2 concentration in such environments is essential in view of reducing the risk of infection of susceptible people exposed. To this end, providing appropriate pathogen-free air supply rates (i.e. air exchange rates) represent a key approach in view of reducing the airborne SARS-CoV-2 concentration in those environments [10,20–22]. Beyond increased dilution, improved ventilation strategies (e.g. personal, displacement) are also needed that can more effectively remove airborne contaminants from the breathing zone, instead of simply dispersing particles throughout the room. Although public transport microenvironments (trains, airplanes, buses) could provide, at least in principle, air exchange rates defined by technical standards (e.g. as a function of the occupancy), the ventilation rates in private cars are set by the passengers according to their air quality/thermal comfort perception rather than contaminant removal concerns. Indeed, an epidemiological study, recently carried out in Singapore in order to explore the transmission risk factors for COVID-19, recognized a significant risk of transmission among non-household contacts sharing a vehicle with an infected subject [23].

A-priori estimates of the risk of infection in cars can be carried out adopting prospective assessments based on a known/estimated emission of virus-laden particles and then diluting them in the indoor environment through either simplified zero-dimensional approaches or complex 3D transient approaches [21,22,24–28]. Zero-dimensional approaches are based on the simplified hypothesis of complete and instantaneous mixing of the emission to achieve a uniform spatial concentration within the environment; they represent a practical solution to easily obtain rough estimates of concentrations and risks when site-specific information regarding the ventilation, geometry, position of the source, etc. are not available [27,29]. Nonetheless, perfectly mixed conditions (hereinafter referred to as "well-mixed") are unlikely in spaces with high ventilation rates [30,31], thus, such hypothesis is even less accurate for large indoor environments characterized by reduced mixing [32] or small confined spaces, as car cabins, where the position of inlet air vents of the HVAC system, the air flow rate entering the car cabin, the air recirculation and its filtration can significantly affect the airflow in the cabin, and then the exposure and risk of passengers [33,34]. Indeed, even if the average air exchange rate in car cabin can be quite high (e.g. >100 $h^{-1}$ when high fan speeds are set or windows are kept open [33,35]), stagnation regions can occur as a result of the specific airflows pattern across the cabin itself. Therefore, for these environments the average risk evaluated through well-mixed models could overestimate or underestimate the actual risk of some of the passengers in the car.

For these reasons, Computational Fluid Dynamics (CFD) represent an essential approach to investigate the risk of infection in cabin cars as it provides detailed information about spatial and temporal virus-laden particle distribution, as a function of specific boundary conditions and ventilation scenarios, by solving the well-known mass, momentum and energy conservation equations alongside with a proper turbulence model [36–42]. To this end, in our previous paper we investigated the particle dispersion in a car cabin through an Eulerian-Lagrangian approach able to perform transient non-isothermal numerical analyses [34]. Such an approach was also experimentally validated against PIV measurements available in the scientific literature [43,44] then providing a validated and suitable approach which can be applied to investigate numerically particle dispersion problems in similar environments.

In this paper we developed and applied an integrated approach aimed at estimating the risk of infection from SARS-CoV-2 Delta variant of susceptible persons sharing the car cabin with an infected person under the outside air intake conditions (i.e. HVAC system in operation, no recirculation, windows closed). The approach here presented integrates a predictive emission-to-risk approach able to determine the risk of infection from the viral load emitted by the infected subject [20,45] with the abovementioned validated CFD approach numerically solved using the open-source OpenFOAM software. The integrated approach was applied to different scenarios to evaluate the effect of the following influence parameters: i) position of the infected subject within the car cabin, ii) air flow rate of the HVAC system, iii) HVAC ventilation mode and iv) expiratory activity (breathing vs. speaking). A further aim of the paper is demonstrating that the risk of infection in indoor environments characterized by fixed seats can be in principle controlled by properly designing the flow patterns of the environment.

## 2 Materials and methods

The integrated approach proposed is based on the following steps:
i)  application of a transient non-isothermal 3D Eulerian-Lagrangian numerical model, developed and validated by the authors in a previous study [34], to describe particle spread once emitted by an infected speaking/breathing passenger located in a car cabin compartment (section 2.1 and 2.2);
ii) description of the emission scenario, i.e. definition of the airborne respiratory particle emission rate of an adult while breathing/speaking [25,28,46] (section 2.3);
iii) calculation of the dose inhaled by the susceptible car occupants for a 30-minute journey through the CFD simulations and estimate of the corresponding SARS-CoV-2 infection risk and number of secondary cases on the basis of the predictive emission-to-risk approach previously developed by the authors [20,45] based on the viral load emitted by the infected subject, the dose of viral load received by the exposed subject, and a dose-response model (section 2.4).

### 2.1 Eulerian-Lagrangian based model to simulate the airborne particle spread within the car cabin

The mathematical-numerical model was developed using the open-source finite volume based OpenFOAM software, to have a fully open and flexible tool with complete control of the variables employed for particle dispersion assessment. It is based on a Eulerian-Lagrangian approach, in which the continuum equations are solved for the air flow (continuous phase) and Newton's equation of motion is solved for each particle (discrete phase).

Velocity, pressure and temperature fields in the car cabin were numerically predicted by solving the mass, momentum and energy conservation equations under the assumption of three-dimensional, unsteady, turbulent and compressible flow with ideal gas behavior. Details about governing Partial Differential Equations (PDEs) are widely available in the scientific literature [47] and are not reported here for brevity.

Turbulence was modelled using the Unsteady Reynolds Averaged Navier Stokes (URANS) approach, and specifically the Shear Stress Transport (SST) $k$–$\omega$ model since the authors in a previous research activity showed that it is the most suitable one to predict airflow patterns within the car cabin under investigation [34]. Details about the employed URANS turbulence model are available in the scientific literature and are not reported here for brevity.

The computed numerical fields were averaged over a selected time interval to reach a quasi-steady state condition: once the quasi-steady state condition is achieved, the flow field is frozen and is used to transport the particles injected by the emitter (i.e. the infected subject) over time during speaking and breathing activities. This approach is exhaustively described in our previous paper [34] to which the interested reader may refer for further information.

The particle motion inside the air flow was modelled by employing the Lagrangian Particle Tracking (LPT) approach, based on a dispersed dilute two-phase flow. In particular, the spacing between particles is sufficiently large and the volume fraction of the particles sufficiently low ($< 10^{-3}$) to justify the use of a Eulerian-Lagrangian approach. The particle motion has been described solving eq. (1) and (2).

$$m_d \frac{d\boldsymbol{u}_d}{dt} = \boldsymbol{F}_D + \boldsymbol{F}_g \tag{1}$$

$$\frac{d\boldsymbol{x}_d}{dt} = \boldsymbol{u}_d \tag{2}$$

where $m_d$ $(kg)$ is the mass of the particle, $\boldsymbol{u}_d$ $\left(\frac{m}{s}\right)$ represents the particle velocity, $t$ $(s)$ is the time, $\boldsymbol{F}_D$ $(N)$ and $\boldsymbol{F}_g$ $(N)$ are, respectively, the drag and gravity forces acting on the particle, $\boldsymbol{x}_d$ $(m)$ represents the trajectory of the particle. The drag force is given by Crowe [48]:

$$\boldsymbol{F}_D = m_d \frac{18}{\rho_d \cdot d_d^2} C_D \frac{Re_d(\boldsymbol{u} - \boldsymbol{u}_d)}{24} \tag{3}$$

In eq. (3), $\rho_d$ $\left(\frac{kg}{m^3}\right)$, $d_d(m)$ and $Re_d$ represent, respectively, the density, diameter and Reynolds number of the particle, $\boldsymbol{u}\left(\frac{m}{s}\right)$ is the air velocity. The particle density was considered constant and equal to 1200 kg m$^{-3}$. The $Re_d$ was calculated as:

$$Re_d = \frac{\rho(|\boldsymbol{u} - \boldsymbol{u}_d|)d_d}{\mu} \tag{4}$$

where $\rho\left(\frac{kg}{m^3}\right)$ is the air density.

The drag coefficient, $C_D$, in eq. (3), is evaluated as a function of the particle Reynolds number as:

$$C_D = \begin{cases} \dfrac{24}{Re_d} & \text{if } Re_d < 1 \\ \dfrac{24}{Re_d}(1 + 0.15 \cdot Re_d^{0.687}) & \text{if } 1 \leq Re_d \leq 1000 \\ 0.44 & \text{if } Re_d > 1000 \end{cases} \tag{5}$$

Particle collisions were considered elastic and the equations of motion for the particles are solved assuming a one-way coupling between the continuum phase and the discrete phase: the flow field affects the particle motion whereas the effect of the particles on the airflow is negligible.

### 2.2 Description of the domain and definition of the boundary conditions and of the scenarios.

The Eulerian-Lagrangian based model, described in Section 2.1, was applied to the analysis of particle dispersion and inhalation in a car cabin evaluating the effects of different geometrical, emission and thermo-fluid-dynamic influence parameters.

The car cabin sizes are 2.47 m × 1.53 m × 1.19 m, corresponding to a total internal volume of 3.46 m$^3$, which is representative of a "large car" according to the United States Environmental Protection Agency (EPA) Fuel Economy Regulations for 1977 and Later Model Year. Four occupants are present in the car cabin and three HVAC system ventilation modes were investigated: front ventilation mode (air entering the cabin through four front vents), windshield defrosting mode (air entering through one vent located under the windshield) and mixed ventilation (all the five vents enabled). All the ventilation modes were tested considering outside air intake provided by a HVAC system, which is with no air recirculation; the windows were considered closed for the entire duration of the journey. Different HVAC flow rates (from 10% to 100% of the maximum flow rate), $Q$ (m$^3$ h$^{-1}$), were investigated: in particular, the intermediate air flow rate (flow rate at the 50% of the maximum fan capacity, hereinafter referred as $Q_{50\%}$) is set at 216 m$^3$ h$^{-1}$ on the basis of the value adopted by Pirouz et al. in their study [49]; this value is consistent with the intermediate fan capacity reported by Ullrich et al. [50] for a real car whose internal volume is comparable to that of the cabin model employed for the scenarios under study. We point out that the flow rate also affects the velocity at the inlet sections and then the velocity field in the car cabin; moreover, in the simulations here performed a single angle of inlet air-flow rate was adopted. Outlet section positions also have an effect on the velocity field in the car cabin, here the outlet sections are placed behind the rear seats in the lower-left and lower-right corners as reported in the experimental case-study adopted to validate the CFD approach here proposed [43,44].

The computational domains employed for numerical simulations are available in Figure 1. The adopted car cabin geometry was built by the authors in a previous study and its description is available in our previous paper [34], together with the explanation of the mesh construction strategy.

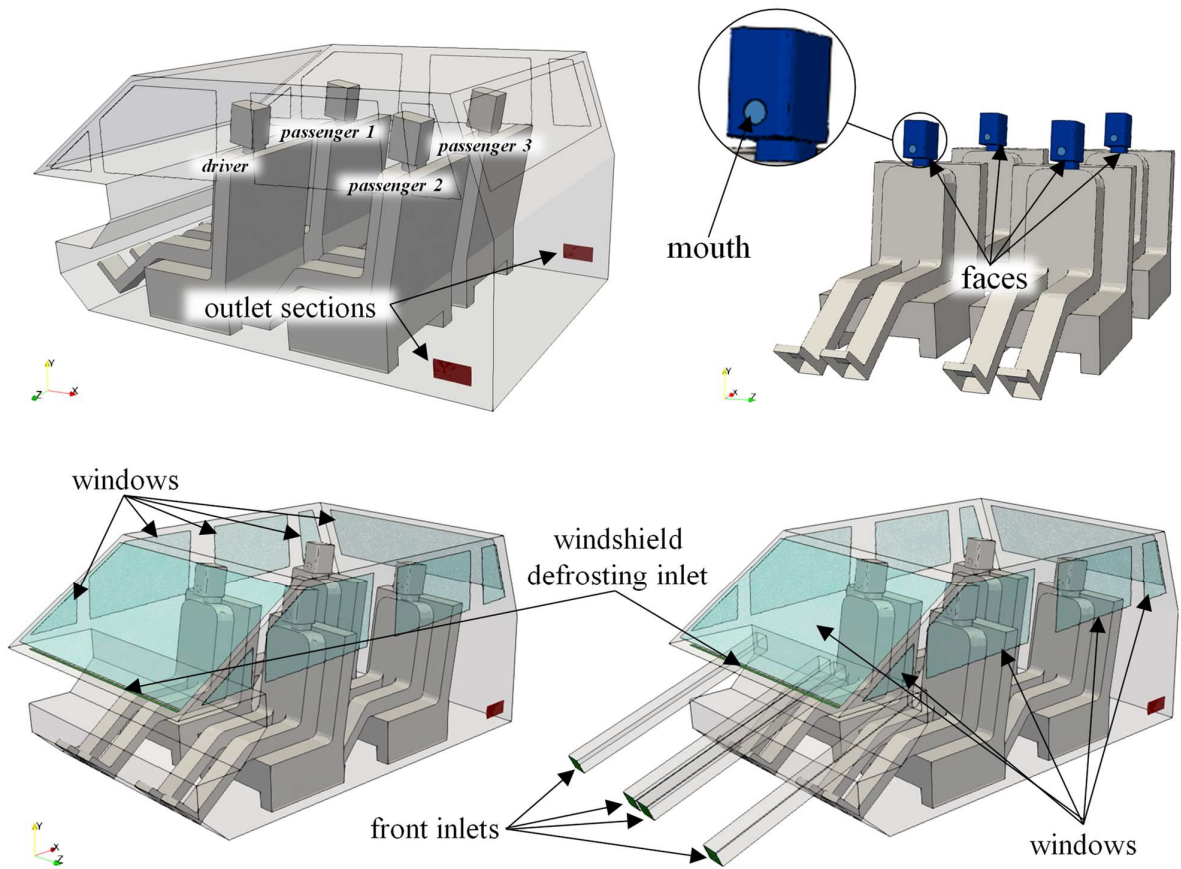

Figure 1 – Computational domain with boundary patches.

In the case of front and mixed ventilation mode, the computational domain is the one in the lower right corner of Figure 1: a rectangular duct is connected to each of the front supply openings, allowing the development of the flow velocity profile before entering the cabin through the air vents; when only front ventilation is enabled, the same computational domain is employed but the patch defining the windshield defrosting inlet is modelled as an adiabatic wall (i.e. the vent is closed). On the other hand, for simulation of scenarios where only windshield defrosting vent is enabled, the computational domain is the one in the lower left corner in Figure 1. Boundaries not specified in Figure 1 have been modelled as adiabatic walls.

In Table 1, Table 2 and Table 3 the boundary conditions imposed for numerical simulations are detailed for the different ventilation modes considered in the present study. Assuming winter climatic conditions, a temperature of 283.15 K was applied to the car windows and inlet air temperature was set to 293.15 K. Passenger face temperatures were set to 306.15 K [51]. Since a relatively high velocity fluid flow was numerically simulated and people in the car cabin were supposed to wear winter clothes with a superficial temperature roughly equal to surrounding air, body temperature plume was neglected.

When both front and windshield defrosting inlets are enabled, the air flow rate is split as follows: each front vent introduces the 12.5% of the prescribed flow rate and the windshield vent the 50% remaining. For turbulent quantities the turbulence intensity $I$ (%) and the turbulent mixing length $\ell(m)$ were specified, the latter calculated as the 7% of the characteristic length $L(m)$ assumed equal to the jet width.

Table 1 - Boundary conditions adopted in case of mixed ventilation (the computational domain employed for such HVAC system operation mode is pictured in the lower right corner in Figure 1).

| surface | BC for velocity | BC for pressure | BC for temperature | BC for $k$ | BC for $\omega$ | Lagrangian |
|---|---|---|---|---|---|---|
| front inlets | (10%) $Q_{10\%} = 43.2$ m³ h⁻¹<br>(25%) $Q_{25\%} = 108$ m³ h⁻¹<br>(50%) $Q_{50\%} = 216$ m³ h⁻¹<br>(75%) $Q_{75\%} = 324$ m³ h⁻¹<br>(100%) $Q_{100\%} = 432$ m³ h⁻¹ | $\frac{\partial p}{\partial n} = 0$ | $T = 293.15\ K$ | $I = 5\%$ | $\ell = 0.07L$ | rebound |
| windshield defrosting inlet | (10%) $Q_{10\%} = 43.2$ m³ h⁻¹<br>(25%) $Q_{25\%} = 108$ m³ h⁻¹<br>(50%) $Q_{50\%} = 216$ m³ h⁻¹<br>(75%) $Q_{75\%} = 324$ m³ h⁻¹<br>(100%) $Q_{100\%} = 432$ m³ h⁻¹ | $\frac{\partial p}{\partial n} = 0$ | $T = 293.15\ K$ | $I = 5\%$ | $\ell = 0.07L$ | rebound |
| outlet sections | $\frac{\partial \boldsymbol{u}}{\partial n} = 0$ | $p = 101325\ Pa$ | $T = 293.15\ K$ | $k = 0.1\ m^2/s^2$ | $\frac{\partial \omega}{\partial n} = 0$ | escape |
| adiabatic walls | $\boldsymbol{u} = 0$ | $\frac{\partial p}{\partial n} = 0$ | $\frac{\partial T}{\partial n} = 0$ | standard wall functions | | escape |
| windows | $\boldsymbol{u} = 0$ | $\frac{\partial p}{\partial n} = 0$ | $T = 283.15\ K$ | standard wall functions | | escape |
| faces | $\boldsymbol{u} = 0$ | $\frac{\partial p}{\partial n} = 0$ | $T = 306.15\ K$ | standard wall functions | | escape |
| emitter mouth | speaking<br>$\|\boldsymbol{u}\| = 1.11\frac{m}{s}$<br>breathing<br>$\|\boldsymbol{u}\| = 0.32\frac{m}{s}$ | $\frac{\partial p}{\partial n} = 0$ | $T = 308.15\ K$ | $k = 0.1\ m^2/s^2$ | $\frac{\partial \omega}{\partial n} = 0$ | rebound |
| receiver mouth | $\|\boldsymbol{u}\| = 0.32\frac{m}{s}$ | $\frac{\partial p}{\partial n} = 0$ | $T = 308.15\ K$ | $k = 0.1\ m^2/s^2$ | $\frac{\partial \omega}{\partial n} = 0$ | escape |

The emitter (infected subject) and receiver (susceptible subject) mouths were modelled as circular surfaces with a radius of 2 cm; a temperature equal to 308.15 K was imposed at the mouths of emitter and receiver. As boundary conditions for air velocity at the mouths of emitter and receiver, fixed velocities equal to 1.11 m/s and 0.32 m/s in magnitude for speaking and breathing activities were respectively imposed as mean values of sinusoidal during exhalation and inhalation reported by Abkarian et al. [52] and Cortellessa et al. [25]. All subjects were assumed mouth-breathers, thus airborne particles were expired and inhaled through the mouth. For particle injection, a random velocity direction from the emitter's mouth, was evaluated as Abkarian et al. [52] considering a conical jet flow with an angle equal to 22°. As concerns the velocity vector direction from the emitter's mouth, a conical jet flow was considered, adopting a cone angle equal to 22° with random velocity directions in intervals of 0.1 s. This adopted angle was calculated by Abkarian et al. [52] to enclose 90% of the particles in a cone passing through the mouth exit and was verified to remain stable with time after the initial cycles. Finally, as concern the boundary conditions of the particles, the Lagrangian Particle Tracking was solved applying an *escape* boundary condition over all the surfaces of the computational domain except for entry sections (for which a *rebound* boundary condition was adopted). In other words, the particles touching the external surfaces (of the domain and of the subjects) disappear and cannot re-enter the computational domain, thus avoiding accumulation of viral load in the environment.

Table 2 - Boundary conditions adopted in case of front ventilation (the computational domain employed for such HVAC system operation mode is pictured in the lower right corner in Figure 1).

| surface | BC for velocity | BC for pressure | BC for temperature | BC for $k$ | BC for $\omega$ | Lagrangian |
|---|---|---|---|---|---|---|
| front inlets | (50%) $Q_{50\%}$ = 216 m³ h⁻¹ | $\frac{\partial p}{\partial n} = 0$ | $T = 293.15\ K$ | $I = 5\%$ | $\ell = 0.07L$ | *rebound* |
| outlet sections | $\frac{\partial \boldsymbol{u}}{\partial n} = 0$ | $p = 101325\ Pa$ | $T = 293.15\ K$ | $k = 0.1\ m^2/s^2$ | $\frac{\partial \omega}{\partial n} = 0$ | *escape* |
| adiabatic walls | $\boldsymbol{u} = 0$ | $\frac{\partial p}{\partial n} = 0$ | $\frac{\partial T}{\partial n} = 0$ | standard wall functions | | *escape* |
| windows | $\boldsymbol{u} = 0$ | $\frac{\partial p}{\partial n} = 0$ | $T = 283.15\ K$ | standard wall functions | | *escape* |
| faces | $\boldsymbol{u} = 0$ | $\frac{\partial p}{\partial n} = 0$ | $T = 306.15\ K$ | standard wall functions | | *escape* |
| emitter mouth | speaking $\|\boldsymbol{u}\| = 1.11\ \frac{m}{s}$ breathing $\|\boldsymbol{u}\| = 0.32\ \frac{m}{s}$ | $\frac{\partial p}{\partial n} = 0$ | $T = 308.15\ K$ | $k = 0.1\ m^2/s^2$ | $\frac{\partial \omega}{\partial n} = 0$ | *rebound* |
| receiver mouth | $\|\boldsymbol{u}\| = 0.32\ \frac{m}{s}$ | $\frac{\partial p}{\partial n} = 0$ | $T = 308.15\ K$ | $k = 0.1\ m^2/s^2$ | $\frac{\partial \omega}{\partial n} = 0$ | *escape* |

Table 3 – Boundary conditions adopted in case of windshield defrosting ventilation (the computational domain employed for such HVAC system operation mode is pictured in the lower left corner in Figure 1).

| surface | BC for velocity | BC for pressure | BC for temperature | BC for $k$ | BC for $\omega$ | Lagrangian |
|---|---|---|---|---|---|---|
| windshield defrosting inlet | (50%) $Q_{50\%}$ = 216 m³ h⁻¹ | $\frac{\partial p}{\partial n} = 0$ | $T = 293.15\ K$ | $I = 5\%$ | $\ell = 0.07L$ | *rebound* |
| outlet sections | $\frac{\partial \boldsymbol{u}}{\partial n} = 0$ | $p = 101325\ Pa$ | $T = 293.15\ K$ | $k = 0.1\ m^2/s^2$ | $\frac{\partial \omega}{\partial n} = 0$ | *escape* |
| adiabatic walls | $\boldsymbol{u} = 0$ | $\frac{\partial p}{\partial n} = 0$ | $\frac{\partial T}{\partial n} = 0$ | standard wall functions | | *escape* |
| windows | $\boldsymbol{u} = 0$ | $\frac{\partial p}{\partial n} = 0$ | $T = 283.15\ K$ | standard wall functions | | *escape* |
| faces | $\boldsymbol{u} = 0$ | $\frac{\partial p}{\partial n} = 0$ | $T = 306.15\ K$ | standard wall functions | | *escape* |
| emitter mouth | speaking $\|\boldsymbol{u}\| = 1.11\ \frac{m}{s}$ breathing $\|\boldsymbol{u}\| = 0.32\ \frac{m}{s}$ | $\frac{\partial p}{\partial n} = 0$ | $T = 308.15\ K$ | $k = 0.1\ m^2/s^2$ | $\frac{\partial \omega}{\partial n} = 0$ | *rebound* |
| receiver mouth | $\|\boldsymbol{u}\| = 0.32\ \frac{m}{s}$ | $\frac{\partial p}{\partial n} = 0$ | $T = 308.15\ K$ | $k = 0.1\ m^2/s^2$ | $\frac{\partial \omega}{\partial n} = 0$ | *escape* |

Careful attention was paid to the computational mesh construction: simulations were performed employing hexahedral-based unstructured computational grids, realized employing the open-source *snappyHexMesh* algorithm. The grid sensitivity analysis was well discussed by Arpino et al. [34] for the same car cabin computational domain and it is not illustrated here for brevity. The adopted grids are composed by 7899968 cells (mixed and front ventilation scenarios) and 7737311 cells (windshield defrosting ventilation scenario) and were properly refined in correspondence of solid walls and in the jet region, where significant velocity gradients are expected. By way of illustration, the computational grid employed for mixed ventilation and front ventilation scenarios is depicted in Figure 2.

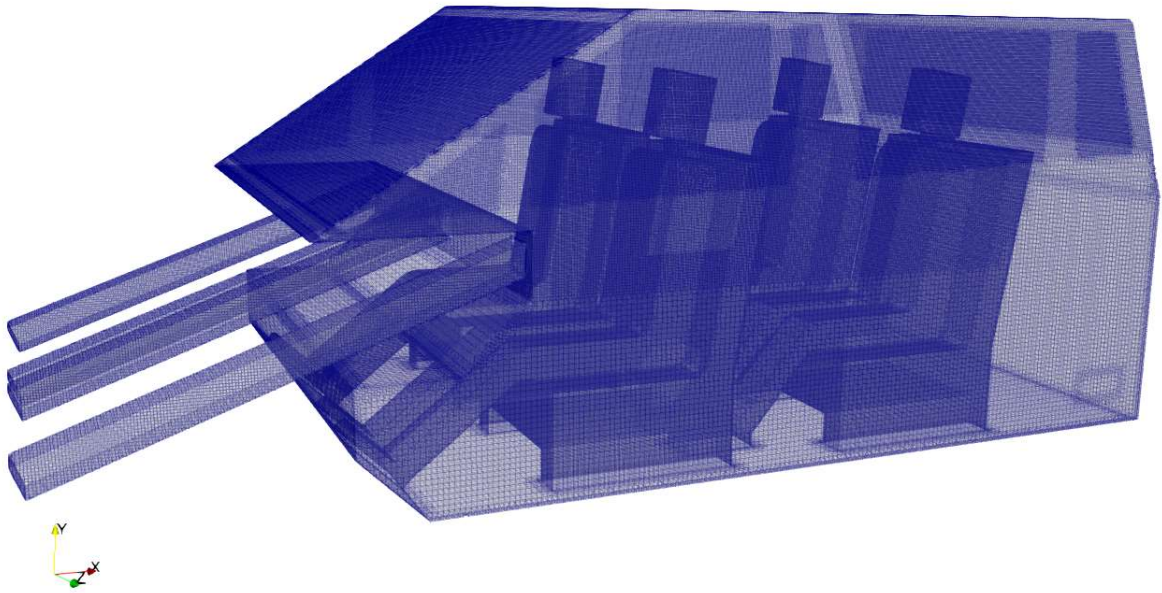

Figure 2 – Computational grid employed for mixed ventilation and front ventilation scenarios.

The risk of infection was evaluated for different exposure scenarios aimed at evaluating the effect of influence parameters under investigation. In particular, the following influence parameters were analysed: i) influence of the position of the infected subject in the car cabin (i.e. driver vs. passenger sitting on the right rear seat), ii) influence of the HVAC system flow rate (i.e. from 10% to 100% of the maximum flow rate, hereinafter also referred as airflow ratio), iii) influence of the HVAC ventilation mode (i.e. mixed, front, and windshield defrosting), iv) influence of the expiratory activity (i.e. breathing vs speaking). The scenarios and the corresponding parameters adopted in the simulations are summarized in Table 4. Please note that the exposed susceptible subjects were considered breathing through the mouth while sitting.

Table 4 – Scenarios investigated through CFD analyses: definition of the parameters adopted to evaluate the effect of the influence parameters.

| Scenarios investigated | position of the infected subject | HVAC system flow rate | HVAC ventilation mode | expiratory activity of the infected subject |
|---|---|---|---|---|
| influence of the position of the infected subject | driver, passenger sitting on the right rear seat (passenger 3 of Figure 1) | $Q_{50\%}$ | mixed | speaking |
| influence of the HVAC system flow rate | driver | $Q_{10\%}$, $Q_{25\%}$, $Q_{50\%}$, $Q_{75\%}$, $Q_{100\%}$ | mixed | speaking |
| influence of the HVAC ventilation mode | driver | $Q_{50\%}$ | mixed, front, windshield defrosting | speaking |
| influence of the expiratory activity | driver | $Q_{50\%}$ | windshield defrosting | speaking, breathing |

## 2.3 Particle emission

Particle emission from the infected subject was modelled as a function of the expiratory activity, i.e. speaking [25] and breathing [46]. In particular, the particle number emission rate ($ER_N$, particle s$^{-1}$), i.e. the size dependent number of particles exhaled by the infected subject per unit time, was estimated for speaking and breathing on the basis of the experimental analyses carried out by Johnson et al. and Morawska et al. [53,54]. Indeed, they measured the number distribution of the particles in the range 0.5-1000 µm in the close proximity of the mouth of an adult person while breathing and speaking. This measurement is extremely complex due to the quick evaporation phenomenon typical of the respiratory particles as soon as they are emitted. For further details on the experimental apparatus and the methodology adopted, the readers are kindly suggested referring to the above-mentioned papers [53,54]. Here, for the sake of brevity, we just show the simplified particle number distributions considered to make the simulations affordable: indeed, we have fitted the original distributions [53,54] through simplified distributions made up of five size ranges. Due to the negligible contribution to the infection risk of sprayborne respiratory particles demonstrated in previous papers [25,55,56], the five size ranges here considered are limited to the airborne respiratory particle range (<90 µm). Volume distributions and emission rates ($ER_V$, µL s$^{-1}$) were calculated considering the particles as sphere. Since the evaporation phenomenon occurs quickly as soon as the particles are emitted [57,58], the post-evaporation number and volume distributions were considered in the CFD model. Indeed, the volume particle distribution before evaporation (i.e. as emitted) was reduced to that resulting from the quick evaporation considering a volume fraction of non-volatiles in the initial particle of 1% [57]. This particle evaporation phenomenon reduces the particle diameter to about 20% of the emitted size. The particle number and volume distributions pre- and post-evaporation (fitted by five size ranges) adopted in the simulations and the corresponding number and volume emission rates are summarized in Table 5 for both the expiratory activities investigated.

Table 5 – Particle number (dN/dlog($d_d$)) and volume (dV/dlog($d_d$)) distributions pre- and post-evaporation fitted by five size ranges as adopted in the simulations for breathing and speaking expiratory activities. Particle number ($ER_N$) and volume ($ER_V$) emission rates are also reported.

| Expiratory activity | Pre-evaporation | | | | | Post-evaporation | | |
|---|---|---|---|---|---|---|---|---|
| | Particle diameter (size range), $d_d$ (µm) | dN/dlog($d_d$) (part. cm$^{-3}$) | dV/dlog($d_d$) (µL cm$^{-3}$) | $ER_N$ (part. s$^{-1}$) | $ER_V$ (µL s$^{-1}$) | Particle diameter (size range), $d_d$ (µm) | dN/dlog($d_d$) (part. cm$^{-3}$) | dV/dlog($d_d$) (µL cm$^{-3}$) |
| breathing | 2.4 µm (<1.9 to 3.2 µm) | 0.312 | 2.33×10$^{-9}$ | 32.6 | 2.43×10$^{-7}$ | 0.5 µm (<0.7 µm) | 0.312 | 2.33×10$^{-11}$ |
| | 4.1 µm (3.2 to 5.4 µm) | 0.016 | 5.80×10$^{-10}$ | 1.6 | 6.11×10$^{-8}$ | 0.9 µm (0.7 to 1.2 µm) | 0.016 | 5.80×10$^{-12}$ |
| | 7.1 µm (5.4 to 9.3 µm) | 0.005 | 1.03×10$^{-9}$ | 0.6 | 1.09×10$^{-7}$ | 1.5 µm (1.2 to 2.0 µm) | 0.005 | 1.03×10$^{-11}$ |
| | 16.0 µm (9.3 to 27.3 µm) | 0.001 | 2.15×10$^{-9}$ | 0.2 | 4.52×10$^{-7}$ | 3.4 µm (2.0 to 5.9 µm) | 0.001 | 2.15×10$^{-11}$ |
| | 35.8 µm (27.3 to 46.9 µm) | <0.001 | 3.44×10$^{-13}$ | <0.1 | 3.63×10$^{-11}$ | 7.7 µm (5.9 to 10.1 µm) | <0.001 | 3.44×10$^{-15}$ |
| | Total | 0.078 | 1.92×10$^{-9}$ | 33.4 | 8.65×10$^{-7}$ | Total | 0.078 | 1.92×10$^{-11}$ |
| speaking | 4.6 µm (< 0.5 to 4.6 µm) | 0.266 | 1.39×10$^{-8}$ | 217.6 | 1.14×10$^{-5}$ | 1 µm (< 1 µm) | 0.266 | 1.39×10$^{-10}$ |
| | 9.0 µm (4.6 to 17.7 µm) | 0.035 | 1.33×10$^{-8}$ | 20.3 | 7.81×10$^{-6}$ | 1.9 µm (1.0 to 3.8 µm) | 0.035 | 1.33×10$^{-10}$ |
| | 23.2 µm (17.7 to 30.4 µm) | 0.013 | 8.75×10$^{-8}$ | 3.1 | 2.05×10$^{-5}$ | 5 µm (3.8 to 6.6 µm) | 0.013 | 8.75×10$^{-10}$ |
| | 45.5 µm (30.4 to 68.2 µm) | 0.016 | 8.08×10$^{-7}$ | 5.7 | 2.83×10$^{-4}$ | 9.8 µm (6.6 to 14.7 µm) | 0.016 | 8.08×10$^{-9}$ |
| | 78 µm (68 to 90 µm) | 0.015 | 3.83×10$^{-6}$ | 1.8 | 4.48×10$^{-4}$ | 16.8 µm (14.7 to 19.2 µm) | 0.015 | 3.83×10$^{-8}$ |
| | Total | 0.249 | 7.71×10$^{-7}$ | 248.6 | 7.71×10$^{-4}$ | Total | 0.249 | 7.71×10$^{-9}$ |

## 2.4 Estimation of the dose received by the susceptible subjects and infectious risk assessment

The risk of infection of the exposed subjects can be calculated based on the dose of viral load (RNA copies considering the viable/infectious fraction) received by susceptible subjects as they inhale virus-laden respiratory particles emitted by the infected subject. Then, a dose-response model is adopted to convert the dose of viral load into a risk of infection.

The dose of viral load is the product of the respiratory particle dose received by the susceptible subjects during the exposure event and the viral load carried by the airborne respiratory droplets emitted by the infected subject. The viral load ($c_v$) carried by the particles was retrieved from data currently available from the scientific literature. Here we adopted the viral load of the Delta variant (B.1.617.2 SARS-CoV-2) which is dominant across much of the world at the time of writing. In particular, we fit the $c_v$ distribution data provided by Teyssou et al. [59] (median value of 7.83 $\log_{10}$ RNA copies mL$^{-1}$) with a quartile simulation approach, in other words we performed a Monte Carlo simulation using proportionally selecting random values within each quartile (1$^{st}$ quartile range 4.3-6.3 $\log_{10}$ RNA copies mL$^{-1}$, 2$^{nd}$ quartile range 6.3-7.8 $\log_{10}$ RNA copies mL$^{-1}$, 3$^{rd}$ quartile range 7.8-8.8 $\log_{10}$ RNA copies mL$^{-1}$, 4$^{th}$ quartile range 8.8-9.4 $\log_{10}$ RNA copies mL$^{-1}$) to obtain 20 000 $c_v$ values. The respiratory particle dose received by each susceptible subject in the car cabin is evaluated through both CFD analyses and well-mixed approach.

### 2.4.1 CFD analyses

Thus, the dose of RNA copies carried by respiratory airborne particles and then inhaled by the susceptible subject for each $c_v$ value ($D(c_v)$) was calculated as

$$D(c_v) = c_v \int_0^T V_{p-pre}(t) dt \qquad \text{(RNA copies)} \qquad (6)$$

where $V_{p-pre}(t)$ is the doses of airborne particles inhaled as a function of the exposure time ($t$), and $T$ is the total exposure time (here considered equal to 30 min). We highlight that the viral load carried by the respiratory particle is related to the initial particle volume (i.e. before evaporation) since the evaporation leads to a reduction in the particle volume (the RNA copies do not evaporate); thus, the $V_{p-pre}$ term has been adopted as the dose of airborne particles calculated with the initial (pre-evaporation) volume. On the contrary, the actual dose in terms of volume of respiratory particles inhaled by the susceptible occupants is referred to the actual volume at the time of inhalation (i.e. post-evaporation; hereinafter referred as $V_{p-post}$). Indeed, we highlight that the respiratory particles dynamics is driven by the post-evaporation particle size, whereas the viral load they carry is a function of the pre-evaporation particle size.

From the dose viral load (i.e. the dose of RNA copies), the probability of infection of the exposed subject for each $c_v$ (($P_I(c_v)$) was calculated adopting a well-known exponential dose-response model [29,60]:

$$P_I(c_v) = 1 - e^{-\frac{D(c_v)}{HID_{63}}} \qquad (\%) \qquad (7)$$

where $HID_{63}$ represents the human infectious dose for 63% of susceptible subjects, i.e. the number of RNA copies needed to initiate the infection with a probability of 63%. For SARS-CoV-2, a $HID_{63}$ value of 7×10$^2$ RNA copies was applied as recently estimated by Gale [61]. We note that in subsequent work Gale [62] increased the RNA copy-to-plaque-forming unit (pfu) ratio used in the thermodynamic dose-response model from 3.6×10$^2$ (based on Vicenzi et al. [63]) to 10$^4$ RNA copies:pfu, which improves agreement with the dose-response estimates of Zhang and Wang [64]. This adjustment increases the $HID_{63}$ value approximately thirtyfold to 2×10$^4$ RNA copies [62]. However, using the golden Syrian hamster model, Hawks et al. [65] found RNA levels in air samples to be ~200 times higher than pfu levels one and two days postinoculation, with infectious virus non-detect afterwards despite the persistence of RNA detections. This indicates a kinetic aspect to the RNA:pfu ratio likely associated with the immune response that affects the infectious virus fraction [66]. As we are focused on modeling the early time period of infection for an infected host previously naïve to SARS-CoV-2, we maintain use of the $HID_{63}$ value of 7×10$^2$ RNA copies herein, which is also generally consistent with the predictions of a novel dose-response approach developed by Henriques et al. [66]. Furthermore, variants of concern such as Delta and Omicron may have greater infectiousness with a lower $HID_{63}$, providing another reason to continue with the original model of Gale [61] given the great uncertainty in the dose-response model for humans.

In order to consider the range of possible viral load values, the individual risk of infection ($R$) of each exposed passenger was calculated through a Monte Carlo simulation with 20 000 realizations, in which the viral load ($c_v$) was sampled randomly from the previously defined distribution and then assigned as the RNA concentration of the exhaled particle volume to calculate the inhaled dose of RNA copies ($D$) and resulting probability of infection ($P_I$) for each realization. The mean of the 20 000 $P_I$ values is calculated as the individual risk ($R$) for each passenger based on their respective inhaled doses.

### 2.4.2 Well-mixed approach

To compare the risks of infection obtained through the detailed CFD analyses proposed here with to this we would have calculated adopting the well-mixed hypothesis, the risk of infection of the susceptible subjects was also assessed adopting the simplified zero-dimensional model assuming complete and instantaneous mixing of the viral emission. In this case the dose of RNA copies received by the susceptibles was estimated on the basis of the average well-mixed viral load concentration in the car cabin $C_{vl,avg}$ (RNA copies m$^{-3}$) over the course of the 30-minute journey, based on the analytical solution of Miller et al. [26]:

$$C_{vl,avg}(c_v, t) = \frac{E_{vl}(c_v)}{V_{cabin} \cdot IVRR} \left[1 - \frac{1}{IVRR \cdot t}(1 - e^{-IVRR \cdot t})\right] \quad \text{(RNA copies m}^{-3}\text{)} \quad (9)$$

where $V_{cabin}$ (m$^3$) is the volume of the car cabin under investigation, $IVRR$ (h$^{-1}$) represents the infectious virus removal rate in the space investigated, and $E_{vl}$ is the viral load emission rate (RNA copies h$^{-1}$). $IVRR$ is the sum of three contributions [67]: the particle deposition on surfaces ($k$, here assumed equal to 0.24 h$^{-1}$ [68]), the viral inactivation ($\lambda$, here assumed equal to 0.63 h$^{-1}$ [69]), and the average air exchange rate via ventilation (AER, h$^{-1}$). The latter was calculated as the ratio between the airflow rate provided by the HVAC systems and the cabin volume: AERs were equal to 12.5, 31.2, 62.4, 93.6, and 124.9 h$^{-1}$ at $Q_{10\%}$, $Q_{25\%}$, $Q_{50\%}$, $Q_{75\%}$, and $Q_{100\%}$ flow rates, respectively. $E_{vl}$ was calculated as the product of the viral load ($c_v$, obtained from simulation as described previously) and the cumulative, pre-evaporation airborne volume emission rate (ER$_V$) obtained from Table 5 (i.e. 7.71×10$^{-4}$ and 8.65×10$^{-7}$ μL s$^{-1}$ for speaking and breathing, respectively.

The dose of RNA copies inhaled by the exposed subject was then estimated as:

$$D(c_v, t) = IR \cdot C_{vl,avg}(c_v, t) \cdot t \quad \text{(RNA copies)} \quad (10)$$

with $IR$ being the inhalation rate and assumed to be 0.54 m$^3$ h$^{-1}$.

As with the analysis based on the CFD results, a Monte Carlo simulation was performed to estimate the individual risk ($R$) of each susceptible passenger based on the viral load of the emitting host. For speaking, a simulation to calculate $R$ using eq. (7) was performed for each of the three ER$_v$ values presented earlier. The distribution of secondary cases and $R_{event}$ were calculated using the Bernoulli trial approach (eq. [8]) for the most representative well-mixed scenarios as further described in Section 3.

In terms of the emission rate in units of infectious doses of Delta SARS-CoV-2, or "quanta" when considering the $HID_{63}$, the equivalent values modeled herein for the 25$^{th}$, 50$^{th}$, and 75$^{th}$ percentile viral loads for speaking are 8.0, 252, and 2524 quanta h$^{-1}$ for the pre-evaporation volume up to 90 μm in diameter. There are no literature values for comparison for the Delta or Omicron variants, but a recent Omicron outbreak at a party in a restaurant in Norway [70] suggests high emission rates are likely. For example, using eqs. (7), (9), and (10) for a ca. 145 m$^2$ room with 3 m ceilings and a 74% probability of infection for a 4.5-hour exposure leads to emission rate estimates of 470 and 1650 quanta h$^{-1}$ for $IVRR$ values of 1.5 and 6.0 h$^{-1}$, respectively, using an $IR$ of ~0.5 m$^3$ h$^{-1}$. Thus, the emission rates evaluated herein appear plausible also considering the rapid spread of both Delta and Omicron variants.

### 2.4.3 Probability of secondary transmission

Beyond the individual risk, which is the mean of an overdispersed distribution and thus masks substantial variability in outcomes, it is of interest to calculate the probability of secondary transmission from the car journey, which is a function of the number of susceptible occupants of the car (*S*). Specifically, the probability of discrete numbers of secondary cases (*C*) arising can be estimated using a Bernoulli trial approach, which is an improvement over past works [21,28] using the percentile values of a continuous distribution of *C* obtained from the simple product of *R* and *S* for each realization. Similar to the methodology of Goyal et al. [71] we model successful transmission for each passenger (assuming all passengers are fully susceptible) by drawing a random uniform variable $U(0,1)$ and comparing it with the $P_I$ value for that passenger, with successful transmission occurring when $U(0,1) < P_I$. This was performed for each of the three susceptible passengers for each realization, and the number of secondary cases (*C*) for an individual realization was calculated by summing up the successful trials as follows:

$$C = \sum_{S=1}^{S=3} Ber(P_I)_S \qquad \text{(secondary cases)} \qquad (8)$$

The end result of the simulation is a discrete probability distribution of secondary cases (*C*), with the mean value representing the event reproduction number ($R_{event}$) of the 30-minute car journey in accordance with the definition of Tupper et al. [72].

## 3 Results and discussion

### 3.1 Influence of the position of the infected subject in the car cabin

Table 6 presents the results of doses in terms of volume of airborne respiratory particle inhaled ($V_{p\text{-}post}$) by susceptible occupants of the car cabin and their individual infection risk for different position of the infected subject (driver vs. passenger #3) in case of mixed ventilation at 50% of the maximum HVAC flow rate ($Q_{50\%}$), speaking activity and 30-minute exposure scenario. Individual risks evaluated through the analytical, zero-dimension well-mixed approach are also reported.

Results show that, in the case of driver infected, the highest dose ($8.68 \times 10^{-9}$ mL) and individual risk (26%) are received by the passenger #2 (left rear seat, i.e. just behind the driver), whereas the passenger #1 (front right seat, i.e. just on the right side of the driver) receives the lowest dose ($1.89 \times 10^{-9}$ mL) and risk (9.2%). Lower doses and risks are received by when the infected passenger #3: the highest dose ($1.42 \times 10^{-9}$ mL) and individual risk (7.2%) are received by the passenger #1, whereas risks lower than 1% are received by the driver and the passenger #2.

Table 6 – Doses in terms of volume of airborne respiratory particle ($V_{p\text{-}post}$) inhaled by susceptible occupants of the car cabin and their individual infection risk for different position of the infected subject (driver vs. passenger #3) in case of mixed ventilation at $Q_{50\%}$, speaking activity and 30-minute exposure scenario. Infection risks evaluated through the well-mixed approach are also reported.

| | Driver infected | | | | Passenger #3 infected | | |
|---|---|---|---|---|---|---|---|
| Susceptible subject | Inhaled volume (mL) | Individual infection risk (%) | | Susceptible subject | Inhaled volume (mL) | Individual infection risk (%) | |
| | | CFD | Well-mixed | | | CFD | Well-mixed |
| Driver | emitter | | | Driver | $5.17 \times 10^{-11}$ | 0.30% | |
| Passenger #1 | $1.89 \times 10^{-9}$ | 9.2% | | Passenger #1 | $1.42 \times 10^{-9}$ | 7.2% | 42% |
| Passenger #2 | $8.68 \times 10^{-9}$ | 26% | 42% | Passenger #2 | $1.59 \times 10^{-11}$ | 0.09% | |
| Passenger #3 | $4.49 \times 10^{-9}$ | 18% | | Passenger #3 | emitter | | |

The reason of such different exposure and risk conditions of the susceptible occupants, occurring as a function of the position of the infected subject, is strictly related to the specific airflow pattern in the car cabin. This is graphically reported in Figure 3 and Figure 4 where streamlines and mean velocity contours as well as the spatial distributions of the airborne respiratory particles after 30 min are reported for driver infected scenario. Figure 3 clearly shows that the streamlines move from the vents, carry the respiratory particles emitted by the driver (slightly moving upwards due to the warm buoyant air exhaled), and convey them towards the passenger just sitting behind him (passenger #2):

therefore a higher exposure to respiratory particles of the passenger #2 occurs as also shown by the spatial distributions of the airborne respiratory particles (Figure 4).

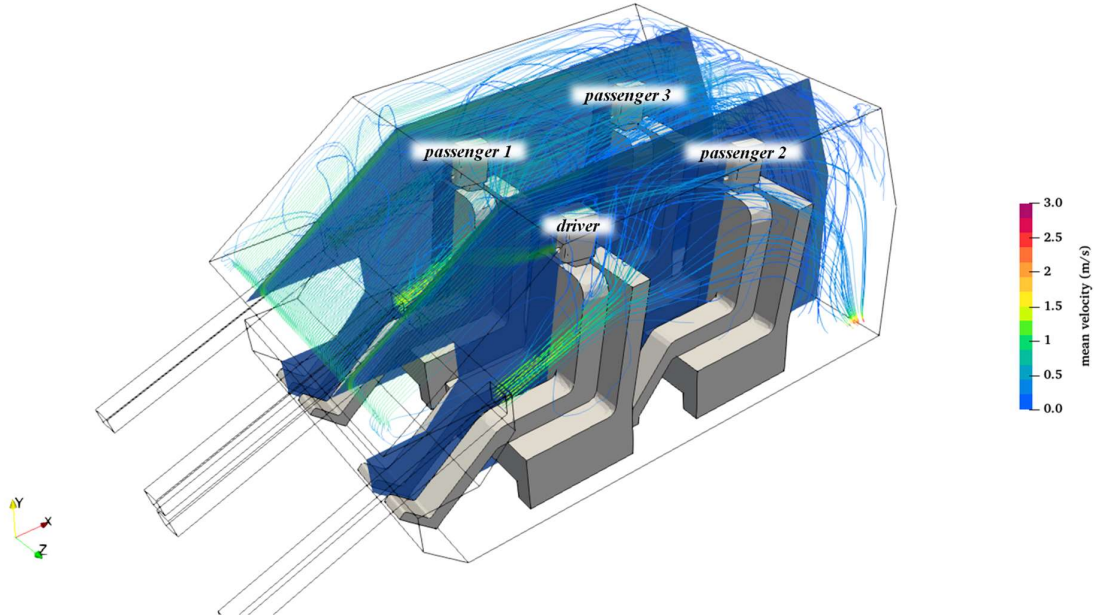

Figure 3 – Streamlines and mean velocity contours on x-y slices at z=-0.38 m and z=0.38 m in case of mixed ventilation mode at 50% ($Q_{50\%}$), speaking activity, driver infected.

A completely different airborne particle distribution can be observed in Figure 5 when the passenger #3 is the infected. In such condition, the airborne particles are mainly confined in the rear seats, but the velocity of particles injected by the infected is sufficient to make them reach the passenger #1, i.e. the passenger sitting just ahead of the infected subject.

For the well-mixed analytical solution, the individual risk is 42% for all passengers regardless of position. This value overestimates the risks received by susceptible subjects in the case of driver infected estimated through the CFD (maximum values 26%) and, even more, the one they receive for the case of passenger #3 being infected (maximum risk 7.2%). The overestimation resulting from the well-mixed approach demonstrates the effectiveness of the HVAC system in reducing the exposure of passengers to virus-laden particles through flow patterns allowing a cleaner air in their breathing zones.

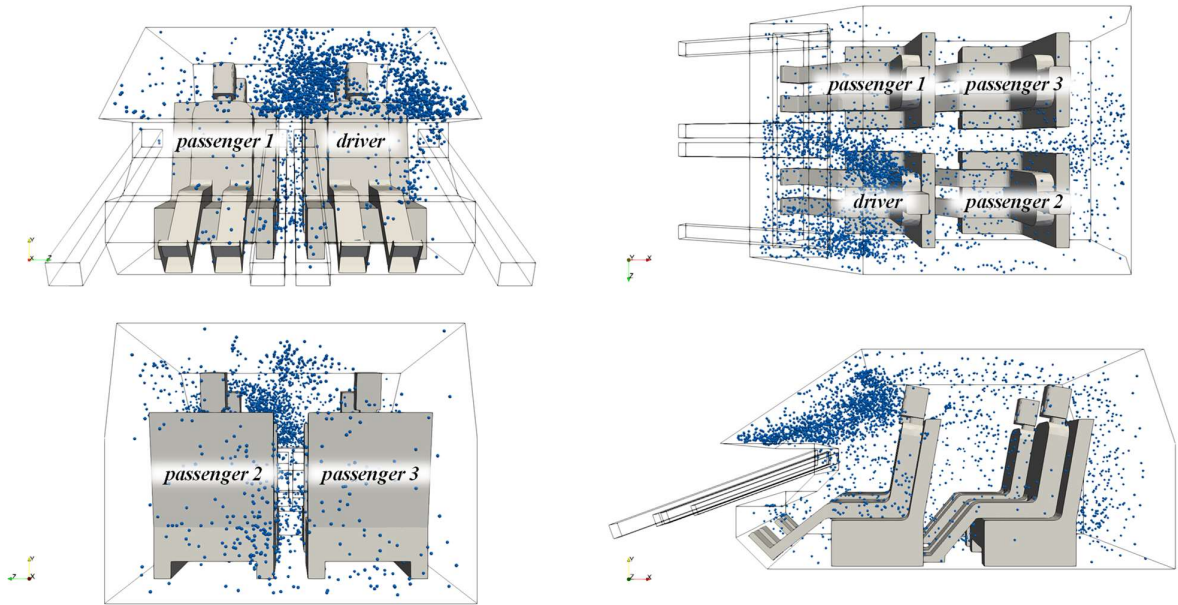

Figure 4 – Spatial particle distribution after 30 min in case of mixed ventilation mode at 50%, speaking activity, driver infected.

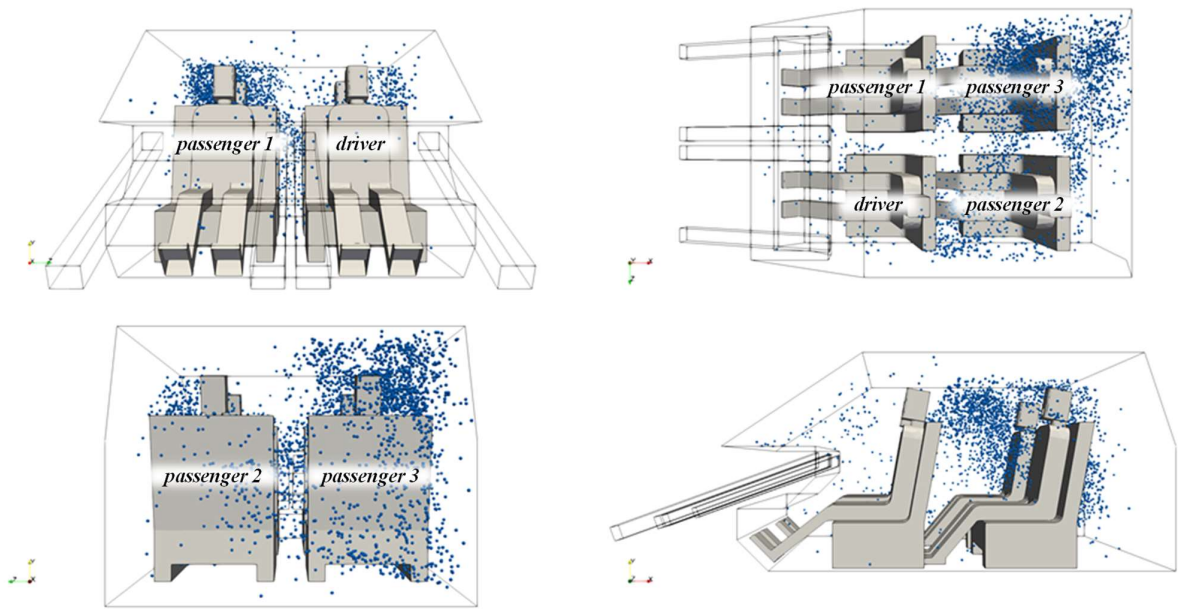

Figure 5 – Spatial particle distribution after 30 min in case of mixed ventilation mode at 50%, speaking activity, passenger #3 infected.

## 3.2 Influence of the HVAC system flow rate

In Table 7 results of doses in terms of volume of airborne respiratory particle ($V_{p\text{-}post}$) inhaled by susceptible occupants of the car cabin and their individual infection risk for different HVAC flow rates (expressed as air flow ratio with respect to the maximum flow rate) in case of mixed ventilation, driver infected, speaking activity, and 30-minute exposure scenario are reported.

When the driver is the infected subject, as already shown in the previous section, the highest doses and risks are (in most of the cases) received by the passenger just sitting behind him/her (passenger #2). As expected, the dose and risk values are strongly influenced by the flow rate provided. As an example, for passenger #2, the risk is <1% for very high flow rates (i.e. $\geq Q_{75\%}$) but it strongly increases with air flow ratios ≤ $Q_{50\%}$ reaching an individual risk of ~50% for $Q_{10\%}$. Similar trends were found for the other passengers with maximum infection risks equal to 32% and 51%, at $Q_{10\%}$, for passengers #1 and #3, respectively. Nonetheless, despite a general decreasing trend of the risk as the HVAC flow rate increases, we point out that the risk of the passengers sitting on the back does not constantly reduce, e.g. the risks at $Q_{25\%}$ and $Q_{75\%}$ are lower than at $Q_{50\%}$ and $Q_{100\%}$, respectively. This is due to the specific air flow patterns occurring at those flow rates which likely undermine the effectiveness of the particle removal towards the exit sections.

In case of low air exchange rate ($Q_{10\%}$) the lowest difference amongst the passengers in terms of risk of infection was detected. This is likely related to the lowest efficiency of the HVAC system in conveying the virus-laden respiratory particles towards the outlet sections, then letting them disperse within the cabin car: indeed, the ratio between the maximum and minimum risk values decreases with the HVAC flow rate then demonstrating a more homogenous concentration. In this respect, it is not surprising that the closest match of the well-mixed results to the average passenger risk calculated through the CFD approach occurs with $Q_{10\%}$ (~12.5 air changes per hour). In this case the CFD-based passenger risks for passenger #1 and #2 is >50% and in good agreement with the well-mixed approach (55%). Conversely, when the air flow ratio is > $Q_{50\%}$, as shown in the previous section, the risk is significantly overestimated using the well-mixed approach.

We point out that all the scenarios here presented consider the HVAC system in operation under the outside air intake conditions; when the HVAC system is not in operation, or it is operated under recirculation ventilation conditions, the actual air exchange rate is clearly lower. Indeed, it is mainly due to the leakages of the car cabin and of the ducts, for this reason, it is strongly affected by the velocity of the vehicles: previous papers showed that the air exchange rate can be lower than 5 h$^{-1}$ [35,73], i.e. well below that obtained under the outside air intake condition at $Q_{10\%}$ flow rate here investigated. For such lower AER values, based on what we have shown above, the well-mixed approach can be considered a useful tool to roughly estimate the risk of exposed subjects: as an example, for a ventilation condition with air recirculation characterized by an AER equal to 2 h$^{-1}$, the estimate of the risk of infection for the passengers provided by the well-mixed approach is >60%.

Table 7 - Doses in terms of volume of airborne respiratory particle ($V_{p\text{-}post}$) inhaled by susceptible occupants of the car cabin and their individual infection risk for different HVAC flow rates ($Q_{10\%}$ to $Q_{100\%}$) in case of mixed ventilation, driver infected, speaking activity, and 30-minute exposure scenario. Infection risks evaluated through the well-mixed approach are also reported.

| HVAC air flow ratio | Inhaled volume (mL) | | | Individual infection risk (%) | | | |
|---|---|---|---|---|---|---|---|
| | Passenger #1 | Passenger #2 | Passenger #3 | Passenger #1 | Passenger #2 | Passenger #3 | All Passengers |
| | | | | CFD | CFD | CFD | Well-mixed |
| $Q_{100\%}$ | 0 | 1.32×10$^{-10}$ | 5.22×10$^{-10}$ | 0 | 0.76% | 2.9% | 35% |
| $Q_{75\%}$ | 4.59×10$^{-12}$ | 7.97×10$^{-11}$ | 3.62×10$^{-10}$ | 0.03% | 0.46% | 2.0% | 38% |
| $Q_{50\%}$ | 1.89×10$^{-9}$ | 8.68×10$^{-9}$ | 4.49×10$^{-9}$ | 9.2% | 26% | 18% | 42% |
| $Q_{25\%}$ | 1.87×10$^{-8}$ | 1.67×10$^{-9}$ | 1.42×10$^{-9}$ | 36% | 8.3% | 7.2% | 48% |
| $Q_{10\%}$ | 8.30×10$^{-8}$ | 1.02×10$^{-7}$ | 1.37×10$^{-8}$ | 51% | 53% | 32% | 55% |

### 3.3 Influence of the HVAC ventilation mode

In Table 8 the doses in terms of volume of airborne respiratory particle ($V_{p\text{-}post}$) inhaled by susceptible occupants of the car cabin and their individual infection risk for different HVAC ventilation mode in case of $Q_{50\%}$ flow rate, driver infected, speaking activity, and 30-minute exposure scenario are reported.

Data clearly highlight that the ventilation mode strongly affect the risk of the passengers. For mixed ventilation mode (air entering the cabin through four front vents), as shown in previous sections, the highest dose is received by the passenger #2 (individual risk 26% at $Q_{50\%}$). Nonetheless, the worst exposure condition is experienced by the passengers for windshield defrosting mode (air entering through one vent located under the windshield) since the risks their passengers range from 22% (passengers #3) to 59% (passenger #2). When a front ventilation mode is adopted the risks of the passengers sitting on the rear seats are almost negligible, whereas the one received by the passenger #1 is extremely high (53%).

Table 8 - Doses in terms of volume of airborne respiratory particle ($V_{p\text{-}post}$) inhaled by susceptible occupants of the car cabin and their individual infection risk for different HVAC ventilation mode in case of $Q_{50\%}$ flow rate, driver infected, speaking activity, and 30-minute exposure scenario. Infection risks evaluated through the well-mixed approach are also reported.

| HVAC ventilation mode | Inhaled volume (mL) | | | Individual infection risk (%) | | | |
|---|---|---|---|---|---|---|---|
| | Passenger #1 | Passenger #2 | Passenger #3 | Passenger #1 | Passenger #2 | Passenger #3 | All Passengers |
| | | | | CFD | CFD | CFD | Well-mixed |
| Front mode | $1.13\times10^{-7}$ | $2.99\times10^{-11}$ | $9.74\times10^{-12}$ | 53% | 0.17% | 0.06% | |
| Windshield defrosting mode | $1.36\times10^{-8}$ | $2.29\times10^{-7}$ | $6.31\times10^{-9}$ | 32% | 59% | 22% | 42% |
| Mixed mode | $1.89\times10^{-9}$ | $8.68\times10^{-9}$ | $4.49\times10^{-9}$ | 9.2% | 26% | 18% | |

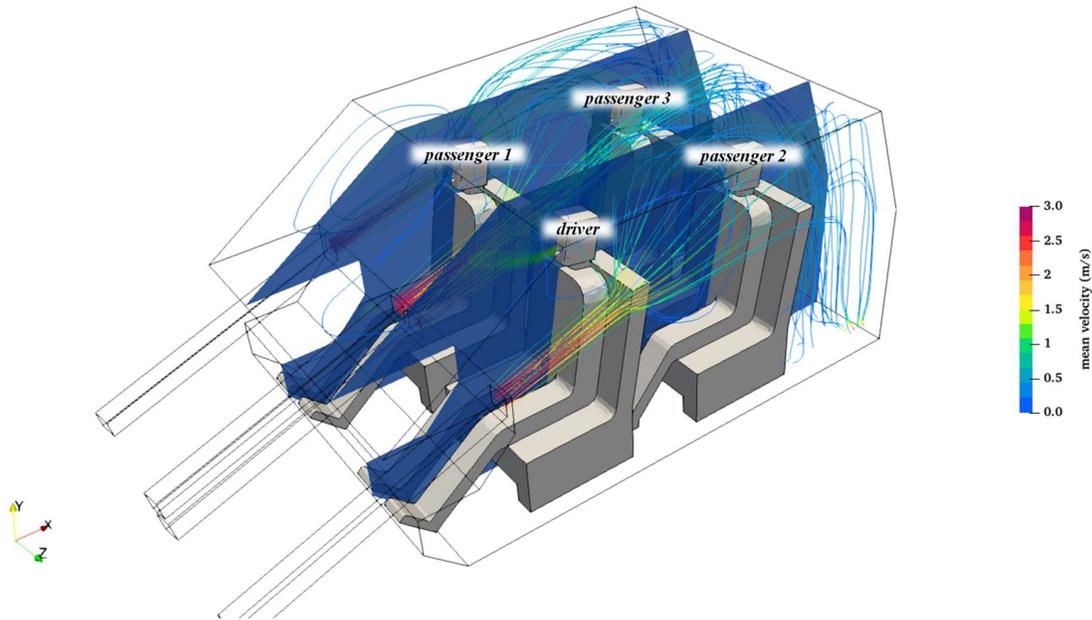

Figure 6 – Streamlines and mean velocity contours on x-y slices at z=-0.38 m and z=0.38 m in case of front ventilation mode at 50% ($Q_{50\%}$), speaking activity, driver infected.

These data can be better explained by referring to the streamlines, main velocity contours and spatial particle distributions. Flow patterns for mixed ventilation mode have been already discussed in the section 3.1, where, for driver infected scenario, the accumulation of respiratory particles in the breathing zone of the passenger #2 has been demonstrated. In case of front ventilation mode, the air flow entering in the cabin impacts the front seats and passengers, changes its direction and forms a recirculation area (Figure 6), then higher concentrations of respiratory particles occur in the front compartment, preventing their spread towards the rear seats (Figure 7) during the whole journey.

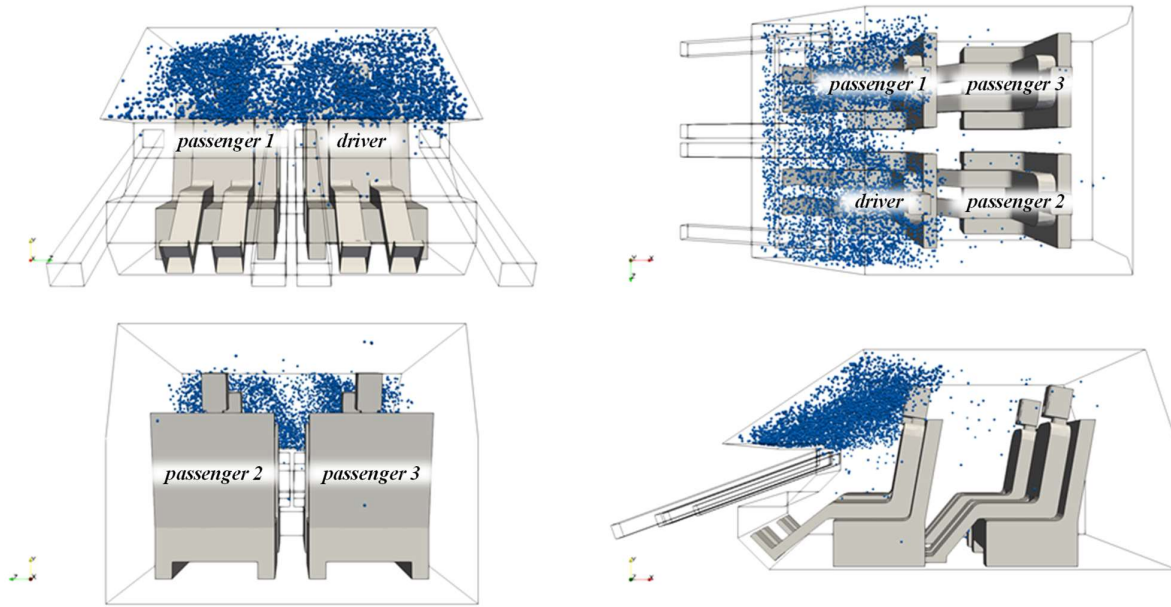

Figure 7 – Spatial particle distribution after 30 min in case of front ventilation mode at 50% ($Q_{50\%}$), speaking activity, driver infected.

On the contrary, in the case of windshield defrosting mode, the respiratory particles emitted by the driver and moving upwards due to the buoyancy forces (Figure 8), are transported in the rear region of the car cabin by the air flow injected through the windshield vent not encountering any obstacle as graphically represented by the streamlines. As a direct consequence of both the airflow patterns and the infected subject position, the airborne particles are mainly confined in the left region of the car (Figure 9) then explaining the reason why the passenger #2 is the most exposed.

Having shown these differences in terms of risk of infection amongst the ventilation modes, it is clear that the well-mixed solution provides a reasonable approximation of the results for the windshield defrosting mode, whereas the front ventilation mode is clearly the least well mixed within the car cabin, and therefore the zero-dimension model significantly overestimates the risk for the back seat passengers by over two orders of magnitude.

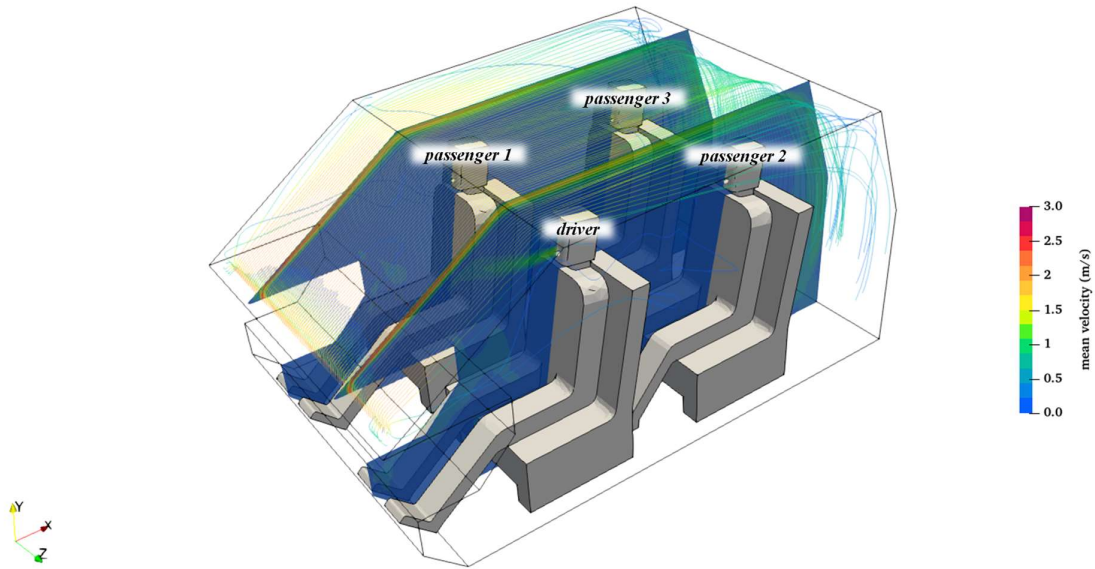

Figure 8 – Streamlines and mean velocity contours on x-y slices at z=-0.38 m and z=0.38 m in case of windshield defrosting ventilation mode at 50% ($Q_{50\%}$), speaking activity, driver infected.

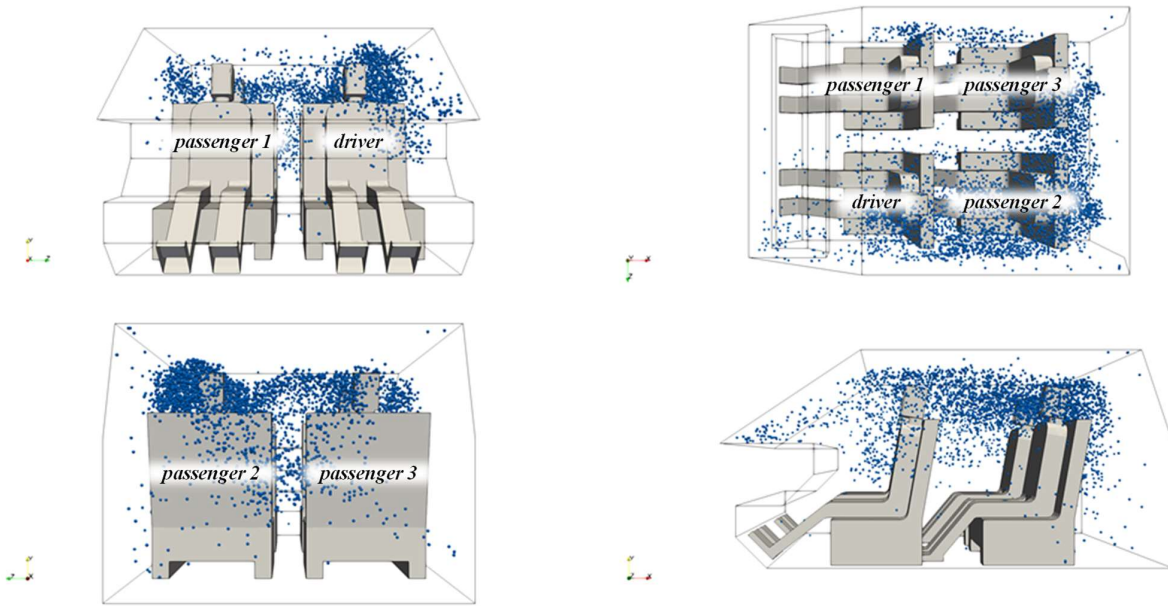

Figure 9 - Spatial particle distribution after 30 min in case of windshield defrosting ventilation mode at 50% ($Q_{50\%}$), speaking activity, driver infected.

### 3.4 Influence of the expiratory activity: breathing vs. speaking

In Table 9 the doses in terms of volume of airborne respiratory particle ($V_{p\text{-}post}$) inhaled by susceptible occupants of the car cabin and their individual infection risk are compared for the two expiratory activities (breathing and speaking) in case of $Q_{50\%}$ flow rate, windshield defrosting ventilation mode, driver infected, and 30-minute exposure scenario (which represents the worst exposure condition amongst those reported in previous section). In the case of breathing, very low airborne particle volumes are inhaled by all the passengers leading to negligible risks of infection (well below 1%): this is due to the low amount of particles emitted and their reduced velocity at the exit of the infected subject's mouth (please see the emission rate discussed in Section 2.3). In the case of breathing activity of the infected subject, the most exposed susceptible is the passenger #3 (not passenger #2 as resulting from speaking activity) and his/her risk (although negligible) is ten-fold the one received by the other two passengers. The difference amongst speaking and breathing activities can also be visually observed comparing the spatial particle distributions of Figure 9 (speaking activity) and Figure 10 (breathing activity) where the latter clearly shows a much lower particle concentration in the car cabin. For the case of breathing, as already reported for speaking, the well-mixed analytical solution provides a rough estimate of the average passenger risk (~0.2% versus ~0.07%).

Table 9 - Doses in terms of volume of airborne respiratory particle ($V_{p\text{-}post}$) inhaled by susceptible occupants of the car cabin and their individual infection risk for different expiratory activities (breathing and speaking) in case of $Q_{50\%}$ flow rate, windshield defrosting ventilation mode, driver infected, and 30-minute exposure scenario. Infection risks evaluated through the well-mixed approach are also reported.

| Expiratory Activity | Inhaled volume (mL) | | | Individual infection risk (%) | | | |
|---|---|---|---|---|---|---|---|
| | Passenger #1 | Passenger #2 | Passenger #3 | Passenger #1 | Passenger #2 | Passenger #3 | All Passengers |
| | | | | CFD | CFD | CFD | Well-mixed |
| Breathing | 2.53×10⁻¹² | 2.18×10⁻¹² | 3.06×10⁻¹¹ | 0.01% | 0.01% | 0.18% | 0.21% |
| Speaking | 1.36×10⁻⁸ | 2.29×10⁻⁷ | 6.31×10⁻⁹ | 32% | 59% | 22% | 42% |

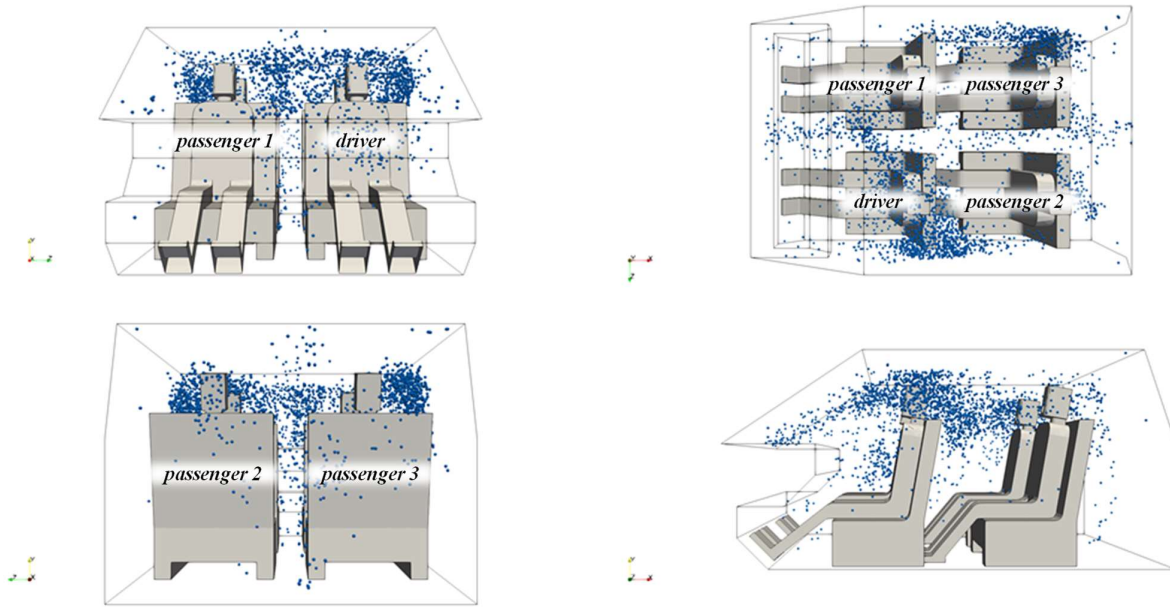

Figure 10 – Spatial particle distribution after 30 min in case of windshield defrosting ventilation mode at 50% ($Q_{50\%}$), breathing activity, driver infected.

### 3.5 Distribution of Secondary Cases

Results of the Bernoulli trial calculations reporting the probability of discrete numbers of secondary cases and the $R_{event}$ are summarized in Table 10 for different scenarios. In particular, in the table all the scenarios tested through the CFD approach are reported as well as the two scenarios presenting well-mixed results comparable to the CFD ones, i.e mixed mode ventilation at $Q_{10\%}$ flow rate for speaking and mixed mode ventilation at $Q_{50\%}$ flow rate for breathing.

The Bernoulli trial data show that there are three model scenarios where the average number of secondary cases ($R_{event}$) exceeds 1 (the $Q_{10\%}$ flow rate condition for both well-mixed and CFD models, and the windshield defrosting mode at the $Q_{50\%}$ flow rate). Supporting the use of the well-mixed approach for $Q_{10\%}$ flow rate, the distribution of secondary cases ($C$) is also very similar to that obtained from CFD, with the probability of zero cases being ~40% and thus the probability of at least one transmission occurring being ~60%. There are three speaking scenarios for which there is over a 90% probability of nobody being infected ($C = 0$) (mixed mode with driver infected at $Q_{75\%}$ and $Q_{100\%}$, and mixed mode with passenger infected at $Q_{50\%}$). For the front mode scenario, there is high risk for the front seat passenger, but the probability that none of the backset passengers gets infected is over 99%. Thus, the front mode is a viable ventilation strategy when the driver is infected and no passenger sits in the front seat, as there is effective aerodynamic containment between the front and back of the car. For the breathing emission rates evaluated herein, there is very low probability of a secondary transmission (~0.2%) for the 30-minute journey.

Table 10 – Results of Bernoulli trial calculations for $R_{event}$ and the probability distribution of secondary cases (*C*) for scenarios under investigation.

| Modeling Scenario | $R_{event}$ | Secondary Case (*C*) Probability | | | |
|---|---|---|---|---|---|
| | | C = 0 | C = 1 | C = 2 | C = 3 |
| Well-mixed approach, $Q_{10\%}$ flow rate | 1.6 | 36.8% | 8.9% | 7.1% | 47.3% |
| CFD mixed mode, driver infected, $Q_{10\%}$ flow rate | 1.3 | 42.0% | 10.9% | 16.9% | 30.2% |
| CFD windshield defrosting mode, driver infected, $Q_{50\%}$ flow rate | 1.1 | 40.4% | 23.7% | 18.4% | 17.5% |
| CFD front mode, driver infected, $Q_{50\%}$ flow rate | 0.54 | 46.5% | 53.3% | 0.26% | 0.00% |
| CFD mixed mode, driver infected, $Q_{50\%}$ flow rate | 0.53 | 66.7% | 17.5% | 11.9% | 3.9% |
| CFD mixed mode, driver infected, $Q_{25\%}$ flow rate | 0.51 | 62.5% | 25.5% | 10.2% | 1.8% |
| CFD mixed mode, passenger infected, $Q_{50\%}$ flow rate | 0.077 | 92.4% | 7.5% | 0.11% | 0.00% |
| CFD mixed mode, driver infected, $Q_{100\%}$ flow rate | 0.036 | 96.3% | 3.7% | 0.06% | 0.00% |
| CFD mixed mode, driver infected, $Q_{75\%}$ flow rate | 0.024 | 97.5% | 2.4% | 0.03% | 0.00% |
| Well-mixed, breathing, $Q_{50\%}$ flow rate | 0.004 | 99.7% | 0.35% | 0.00% | 0.00% |
| CFD windshield defrosting mode, breathing, driver infected, $Q_{50\%}$ flow rate | 0.002 | 99.8% | 0.21% | 0.01% | 0.00% |

### 3.6 Strengths and weaknesses

The results showed in the previous sections highlight the strengths of the CFD approach for a proper evaluation of the risk of infection in small confined spaces affected by a particular fluid dynamics due to high flow rates entering the cabin, or ventilation systems not designed for mixing (e.g. front). Simplified analytical approaches, such as zero-dimensional models, may inaccurately estimate the risk of the exposed subject by a large amount. However, for the 10% flow condition and mixed mode ventilation, the zero-dimension well-mixed approach produces quite similar results in terms of both the average risk (and thus $R_{event}$) and the probability distribution of secondary cases. The parameters under which well-mixed approaches are most defensible requires further evaluation, using CFD and possibly field investigations (e.g. tracer tests) to inform such generalizations; nonetheless, it is clear that well-mixed models can perform very well in scenarios characterized by low air exchange rates where the flow patterns are not able to provide a proper particle removal from the breathing zone of the exposed subject: this is also typical of other larger indoor environments, such as naturally-ventilated buildings [32], where well-mixed models were shown to predict the attack rates of documented SARS-CoV-2 outbreaks [21,27].

We note the solutions here reported are very specific of the cabin car under investigation and of the boundary conditions set. Therefore, we point out that generalizing the obtained CFD results to other passenger vehicles could lead to mistakes too. Indeed, cabin cars comparable in terms of volume and emission rates could present different infection risks for the susceptible occupants as a function of the position of the inlet vents (some cars also have ducts to the rear-seat area), the adjustable angle of inlet air-flow rate, the air flow rate split amongst the different vents, and the position of the outlet sections (considering that in actual cars the particle exfiltration just relies upon leakages of the cabin): these aspects are here not considered and could be involved in future developments of the study. Regardless, our results show that CFD is necessary to evaluate the fate of these particles more accurately and that a proper design of the HVAC system (e.g. in terms of positioning of the inlet and outlet vents, etc.), in view of significantly reducing the risk of infection, is suitable. This is a key finding since it demonstrates that in indoor environments characterized by fixed seats the risk of infection can be in principle controlled by properly designing the flow patterns of the environment, i.e. moving towards an *ad-hoc* personalized ventilation [74,75].

## 4 Conclusions

In the paper we proposed and applied an integrated approach combining a validated CFD transient approach (numerically solved using the open-source software OpenFOAM) and a recently developed predictive emission-to-risk approach in order to estimate the SARS-CoV-2 Delta variant risk of infection in a car cabin under different conditions in terms of ventilation (ventilation mode and air flow rate of the HVAC system) and emission scenarios (expiratory activity, i.e. breathing vs. speaking, and position of the infected subject within the car cabin).

The results of the study clearly showed that the risk of infection, and consequently the distribution of secondary cases, is strongly influenced by the ventilation mode, the HVAC flow rate, the position of the infected subject, and the expiratory activity. As an example, in case of driver infected speaking for the entire journey, a reduced ventilation (low flow rate) or a less effective ventilation (e.g. windshield defrosting mode) can cause high risk of infection then leading to a high probability of at least one secondary case in only 30-min of exposure. The risk of infection is clearly reduced when higher flow rates enter the car cabin then diluting the virus-laden respiratory droplets emitted by the infected subject or when the infected subject just breathes instead of speaking.

CFD approaches are needed to properly address the individual risk in such confined spaces as the fluid-dynamic conditions significantly affect the airflow patterns and the spatial distribution of the virus-laden respiratory particles within the cabin. Thus, simplified zero-dimensional approaches assessing the average risk of the susceptible (not accounting for the specific flow patterns in the confined space), can lead to miscalculation of the risk of the exposed subjects, particularly when ventilation systems are not designed for mixing. Indeed, the well-mixed solutions for speaking infected subject here shown are roughly comparable with the CFD ones only in case of very low flow rates, i.e. when the reduced air flow rates do not effectively clean the breathing zone of the exposed subjects and the virus-laden concentration are likely homogenous within the car cabin. Furthermore, the front ventilation mode evaluated herein provides effective aerodynamic containment between the front and back of the vehicle, meaning passengers sitting in the back seats are better protected from an infected driver relative to mixing ventilation.

Summarizing, CFD modeling is a valuable tool to produce such recommendations for specific applications, which are not possible with simple zero-dimension models and, even if the CFD results here provided are not directly transferable to other cars (due to the case-specific geometry, vent positions, etc.), the finding here indicates that *ad-hoc* designing of the air flow of closed environments in view of reducing and controlling the risk of infection is achievable, especially when the spatial locations of the occupants are fixed.